\documentclass[twocolumn,aps,prd,amsmath,showpacs,nofootinbib]{revtex4}
\usepackage{graphicx} % Include figure files
\usepackage{bm}
\usepackage{txfonts}
\usepackage{mathrsfs}

\def\bq{{\mathbf q}}
\def\br{{\mathbf r}}

\def\b0{{\mathbf 0}}

\def\b0{{\mathbf 0}}

\def\bra{\langle}
\def\ket{\rangle}

\def\alf{\alpha}

\def\Lam{\Lambda}

\def\sg{\sigma}

\begin{document}

\title{Indications of isotropic Lifshitz points in four dimensions}
\author{Dario Zappal\`a}
\affiliation{INFN, Sezione di Catania, Via Santa Sofia 64, 95123 Catania, Italy}
\date{\today}
\begin{abstract}
The presence of isotropic Lifshitz points for a  $U(1)$ symmetric  scalar theory is investigated  with the help of the Functional Renormalization Group at the conjectured lower critical dimension  d=4.
To this aim, a suitable truncation in the expansion of the effective action in powers of the field is considered and, consequently, the Renormalization Group flow is reduced to a set of ordinary differential equations
for the parameters that define the effective action. Within this approximation,  indications of a line of Lifshitz points are found, that present  evident similarities with the properties shown by the line  of fixed points observed 
in the two dimensional Berezinsky-Kosterlitz-Thouless  phase. In particular, this line of Lifshitz points exhibits  the vanishing of the expectation value of the field, together with a finite stiffness and, for specific 
combinations of the parameters that define the effective action,  also the  algebraic decay at large distance of the order parameter correlations.
\end{abstract}

\maketitle

%%%%%%%%%%%%%%%%%%%%%%%%%%%%%%%%%%%%%%%%%%%%%%%%%%%%%%%%%%%%%%%%%%%%%%%%%%

\section{Introduction}

%%%%%%%%%%%%%
Among the various classes of fixed points that characterize the structure of   phase diagrams,  the Lifshitz point  (LP) is of particular interest  because it is associated to the coexistence of three phases which count, together with 
the two, more common, ordered (with an homogeneous  order parameter)  and disordered (with zero order parameter)  phases, an inhomogeneous phase where the order parameter, instead of being  constant, is spatially 
modulated  with finite wave vector\cite{Horn}. 

Around the LP, phase separation lines are determined by the interplay between the derivative term involving  the  square  gradient of  the field $O(\partial^2)$, and the higher derivative operator  $O(\partial^4)$, 
when the coefficient of the former vanishes or becomes negative and its effects are contrasted by the latter, which  instead  becomes the leading derivative term of the action.
The first analysis of the properties of the LP was presented in  \cite{Horn} (see also \cite{erzan,sak,grest}) for the  general case of the anisotropic  LP in which  the space coordinates of the $d$ dimensional space are separated in 
parallel and orthogonal components, respectively  spanning a  $m$ and a $d - m$ dimensional space,  and while the  coordinates of  the $d - m$ dimensional set have standard scaling laws, for the coordinates  belonging to the 
$m$-dimensional subspace, the scaling is regulated by the higher derivative operator $O(\partial^4)$. Therefore the scaling law of each parameter entering the effective action is modified accordingly \cite{Horn,Diehl}.

The LP theory finds application in various phenomenological  contexts such as liquid crystals, or polymer mixtures, or high-Tc superconductors, or magnetic systems (for reviews see \cite{selke1988,Diehl}), 
but also in the high energy sector, LPs turn out to be  relevant in the formulation of emergent gravity  \cite{horava,filippo,zz,cognola}  or higher spin gravity \cite{bekaert1,bekaert2}  or in the  study of  Lorentz
invariance violation\cite{Alexandre,kikuchi}, as well as in the analysis of dense quark  matter\cite{casal2,buballa,pisarski}.

The nature  of the LP  is essentially established by the three parameters ($d,\,m,\,N$), where $N$ indicates the number of fields and, in particular, its properties are crucially related to the parameter of anisotropy $m$. 
So, for instance, the analysis in \cite{Horn} indicates a $m$-dependent  upper  critical dimension $d_u(m)=4+m/2$ while, as indicated in \cite{Diehl}, one expects for the lower critical dimension of the $O(N)$ theory:  
$d_l(m)=2+m/2$.  At the same time it is clear that even the  techniques selected to analyze the problem must be adapted to the specific  value of $m$.  In fact, the difficulties encountered in handling  the propagators 
for generic  $m$ and $d$ (see \cite{parisshpot}),  made the calculation of  the $O(\epsilon^2)$  contribution in the $\epsilon$-expansion a very difficult task which was eventually solved in \cite{carneiro,diehl1,diehl2}.
Analogous difficulties appeared in the computation of the critical properties at large N with the relative O(1/N) corrections \cite{diehl3}. Another  approach  adopted in the analysis  of the anisotropic LPs is  the Functional 
Renormalization Group (FRG)  \cite{Wetterich:1992yh,Morris:1994ie,Berges:2000ew} technique, which has initially been employed to investigate the structure of the uniaxial, $m=1$, LP \cite{bervillier,Essafi}.

Among the various models corresponding to different values of the parameter $m$, the isotropic case, defined by $m=d$, is simpler to treat because the  symmetry of the $d$ coordinates is recovered. This specific case has 
been analyzed  in the $\epsilon$-expansion, by expanding around $d=8$ with  $\epsilon=8-d$ \cite{Horn,erzan,diehl4}, as well as by a numerical Monte-Carlo study \cite{schmid}.  
In addition, more recently,  the FRG was used in the analysis of the isotropic LP,  both for the Ising-like  theory\cite{boza}  (i.e. $N=1$), and for the $O(N)$ symmetric theory\cite{zappa}.  

For the $N=1$ theory, a numerical resolution of the FRG flow equations in the derivative expansion was performed, by resorting to the proper time representation of the flow equations 
\cite{liao,bohr,boza2001,Litpaw3,Litpaw4}, as this scheme already turned out to be quite accurate in the numerical analysis  of the critical  properties of a theory\cite{boza2001,zap2001,Maza,litimzappala} 
and, in addition, the differential flow equation of the  $O(\partial^4)$ operator (necessary in the study of the LP) had been derived before\cite{litimzappala}.  In \cite{boza}, it is shown that beyond the lowest order 
approximation (known as local potential  approximation),  the LP solution is found  in the range $5.5 < d<8$,  with negative anomalous dimension (and the negative sign is confirmed in  the analysis of \cite{safari1}),
but it is not clear whether the lack of solutions at smaller  $d$ is a physical  effect due to the action of the infrared fluctuations, or it is related to some drawbacks in  the numerical analysis.

On the other hand in \cite{zappa}, the flow equations of the  $O(N)$ theory are treated in the framework of the  $1/N$ expansion and the LP solution is observed in the range $4<d<8$, as expected from the general 
expressions of $d_l$ and $d_u$, with the anomalous  dimension $\eta\to 0$ when  $d\to 4_+$. This analysis  suggests that many properties of the isotropic LP of the $O(N)$ theory  between $d_l$ and $d_u$ could 
resemble those of the Wilson-Fisher fixed point between the lower and upper critical dimensions  associated to this point (which respectively are $d=2$ and $d=4$), as, for instance, the fact that the isotropic LP in $d=4$ 
could have a multicritical  nature (with regard to this point, see also \cite{gubser,safari1,safari2}).

Following the above indication, in this paper we explore the possibility that non-trivial properties observed for the scalar theory in $d=2$ dimensions, could have a counterpart related to the LP in $d=4$. 
In fact, although in $d=2$  the Coleman-Mermin-Wagner theorem\cite{mermin66,coleman}  forbids,  for theories with a continuous symmetry, phase transitions  of the kind observed in $d=3$,  a different kind of phase 
transition of topological nature, namely the Berezinski-Kosterliz-Thouless  (BKT) transition \cite{berezinskii71,kosterlitz73}, nevertheless occurs  from a  disordered to a  quasi-ordered phase, for the $O(2)$ symmetric theory.  
Accordingly, it is worth exploring  whether at the lower critical dimension of the isotropic LP, $d_l=4$, a similar non-standard transition could show up.  To this purpose, we resort to the  FRG approach which is particularly suitable for 
the analysis of the scaling  associated to  an isotropic  LP  and, at the same time,  has already been used for reproducing  the main features of the BKT transition.  

In fact the FRG determines,  through  a functional flow  equation, the evolution with the running  scale $k$ of the effective action, starting from the  bare action, given as initial condition of the equation at  some ultraviolet scale 
$k=K_{UV}$, down to  the generator of the  one-particle irreducible vertex functions, which is obtained at $k=0$.
Then,  by using a derivative expansion to parametrize the effective action, the  $O(2)$ theory was studied in $d=2$ \cite{grater, wetterich,dupuis}, by pointing out peculiar properties of the BKT phase transition such as  the essential 
singularity  of the correlation length when the transition is approached from the disordered phase, or the continuous line of fixed points associated with the algebraic decay of the order parameter correlations at large distance
in the BKT phase.  In addition,  the critical value of the anomalous dimension and of the stiffness at the transition point, associated with the Kosterlitz-Thouless temperature, are accurately  reproduced\cite{dupuis}.

In the case of the BKT transition, this approach amounts to solving three coupled partial differential equations for the effective potential and the two coefficients of the $O(\partial^2)$ terms of the effective action, 
with these three variables depending both on the scale $k$ and on the field $\phi$. If extended to the study of the  LP, this approach would require the simultaneous resolution of two additional differential equations, as
in this case the inclusion  of the  $O(\partial^4)$  operators is needed. This turns out to be a rather complex numerical exercise and therefore it would be preferable to rely on a less demanding approximation scheme.

Actually, a simpler treatment of the flow equations, that reproduce at least partially the main properties of the BKT transition, is presented in \cite{jame}, where the effective action is expressed by means of a truncated 
expansion in powers of the field and parametrized in terms of a minimal set of variables that depend on the scale $k$ only. In particular, the coupled partial differential equations are reduced to a set of ordinary differential equations
for the mass of the radial coordinate, the field expectation value and the two wave function renormalization parameters.
 
This simple truncation succeeds in recovering a (approximate) continuous line of fixed points and also  the algebraic decay of the order parameter correlations, thus  providing a  good description of BKT phase (which is realized 
at large values of the stiffness parameter $J$); however it becomes less accurate at smaller $J$ and  fails  in reproducing the transition point. Interestingly, in \cite{jame} it is shown an enlightening comparison between  the results 
obtained  in this truncation, either with the cartesian decomposition of the complex field $\phi$, or by using a modulus-phase representation of $\phi$, that clarifies the origin of some shortcomings of the cartesian decomposition,
and therefore the limits of the approximation considered. This analysis provides the simplest scheme that can be suitably adjusted to study the isotropic LP of a $O(2)$ (or equivalently a $U(1)$) symmetric theory in $d=4$, and 
investigate whether properties similar to those observed  in the BKT phase are recovered also in this case.

The scheme of the paper is the following.
In Sec.~\ref{II} we  examine the general structure of the  $U(1)$ symmetric effective action in the derivative expansion and determine the conditions that lead to  the algebraic, large distance decay of the order parameter correlation
for the isotropic LP case. 
In Sec.~\ref{III} we adapt  the flow equations determined in \cite{jame}, to our problem. In Sec.~\ref{IV} the results obtained by neglecting the longitudinal fluctuations, are presented.
In  Sec.~\ref{Numerical} the numerical output obtained with the inclusion of the longitudinal fluctuations is shown and the conclusions are reported in Sec.~\ref{VI}.

%%%%%%%%%%%%%%%%%%%%%%%%%%%%%%%%%%%%%%%%%%%%%%%%%%%%%%%%%%%%%%%%%%%%%%%%%%
\vskip 8 pt
{\section{Effective action}
\label{II}}

We focus on a four dimensional scalar theory,  whose degrees of freedom are described by a complex field $\phi(\br)$,
with a $U(1)$-invariant effective action, and the FRG  properties are derived from the  general flow equation for the  scale  dependent effective action $\Gamma_k[\phi]$, \cite{Berges:2000ew}, :
\begin{equation} 
\label{rgfloweq}
\partial_t \Gamma_k[\phi]=\frac{1}{2} \int_q\, \partial_t R_k(q)
\left [ \Gamma_k^{(2)}[q,-q;\phi]+R_k(q)\right ]^{-1}
\end{equation} 
that describes the evolution of $\Gamma_k[\phi]$ from the  bare action, taken at some large UV scale, $k=K_{UV}$, down to the 
full effective action, i.e. the generator of the one-particle irreducible vertex functions, obtained at $k=0$.
In Eq.\,(\ref{rgfloweq}), $t\equiv {\rm ln}(K_{UV}/k)$ (and $\partial_t =- k \partial_k$), 
$ \Gamma_k^{(2)}[q,-q;\phi]$ is the second functional derivative of  $\Gamma_k[\phi]$ with respect to 
the field, and  $R_k(q)$  is a suitable  regulator that suppress the modes with $q\ll  k$ and allows to integrate  those with  $q\gg k$.

Then, the main  issue concerns the choice of a specific  parametrization for  $\Gamma_k[\phi]$,  and therefore an approximation that is 
sufficiently comprehensive to exhibit the relevant  properties of the theory. In \cite{jame}, it is shown that many fundamental properties 
of the low temperature BKT phase in $d=2$ dimensions are recovered in this framework, by including very few terms in the corresponding 
parametrization  of the effective action. Namely, besides a mexican hat-like quartic potential, characterized by three parameters (mass,  quartic 
coupling  and field expectation value), the derivative part of the action consists of the two operators $Z\, \left ( \partial \phi(\br) \, \partial \phi^*(\br) \right )$ 
and $Y\,  \left( \partial |\phi(\br)|^2 \right)^2$, where the two parameters $Z$ and $Y$ are field independent.
These two operators correspond to the minimal choice in a derivative expansion of the effective action of a two component $U(1)$-invariant  theory,
in a non-symmetric phase. In fact, the term proportional to $Y$ is essential to recover different renormalization constants for the longitudinal and 
transverse components of the field, despite  it is subdominant with respect to $Z$, due to the higher scaling dimension of the associated operator.

According to these results, when considering the LP in $d=4$ dimensions, where the main derivative terms are operators containing four 
derivatives of the field, one can simply consider the straightforward generalization of the choice made in \cite{jame}, by taking the two operators 
$W_A\, \left ( \partial^2 \phi(\br) \, \partial^2 \phi^*(\br) \right )$  and $W_B\,  \left( \partial^2 |\phi(\br)|^2 \right)^2$, 
again with different  normalization of the longitudinal and of the transverse component of the field $\phi$.
However, one must be aware that a complete treatment of the derivative expansion at the fourth order, respecting the $U(1)$ symmetry,
includes many more terms. For instance, we consider  the  effective action  $\Gamma^E_k[\phi] $  that includes, in addition to the potential $V_k (\rho (\br) )$ 
(with $\rho (\br) \equiv  |\phi(\br)|^2$), operators containing up to the fourth power of the field and up to four field derivatives,
\begin{eqnarray} \label{toym}
\Gamma^E_k[\phi] &=&\int d^4\br  \Bigg\{  V_k (\rho ) +  \left (a_1+a_2 \rho \right ) 
\, \left (\partial \phi \partial \phi^* \right ) + a_3 \left[ \phi^* \overleftrightarrow {\partial} \phi \right]^2
\nonumber \\
&+& \left (b_1+b_2 \rho \right ) \,  \left [ \partial^2 \phi \, \partial^2  \phi^* \right ] + b_3  \left[ \phi^* \overleftrightarrow {\partial^2} \phi \right ]^2 \nonumber\\
&+& b_4 \, \partial \phi \, \partial \phi^* \left[ \phi^* \overleftrightarrow {\partial^2} \phi\right] 
\nonumber \\
&+& b_5  \, \Bigl [\left(\partial \phi \partial \phi \right)\left( \phi^* \partial^2 \phi^* \right ) +\left(\partial \phi^* \partial \phi^* \right) \left( \phi \partial^2 \phi\right) \Bigr]
\nonumber\\
&+& b_6 \, \Bigl [\partial \phi \partial \phi^* \Bigr ] ^2 +b_{\,7} \, \Bigl [\left( \partial \phi \partial \phi \right)\left(  \partial \phi^* \partial \phi^* \right)\Bigr ]  \Bigg\} \; ,
\end{eqnarray}
where the $k$-dependent parameters $a_i$ and $b_i$ are respectively associated to the $O(\partial^2)$ and $O(\partial^4)$ operators,
with $a_1$ and $b_1$ corresponding to terms that are quadratic  in the field, and where we defined 
$\Bigl[ \phi^* \overleftrightarrow {\partial} \phi\Bigr] \equiv \phi^*  \partial \phi + \phi \partial \phi^*$, as well as
$\Bigl[ \phi^* \overleftrightarrow {\partial^2} \phi\Bigr] \equiv \phi^*  \partial^2 \phi + \phi \partial^2 \phi^*$.  
One can easily identify the coefficient $Z$ and $Y$ of  \cite{jame} respectively with $a_1$ and  $a_3$ of Eq.\,(\ref{toym}), 
while $W_A$, introduced above, corresponds to $b_1$ and $W_B$ to a combination of $b_3,\,b_4$ and $b_6$.

For  $V_k (\rho)$ in Eq.\,(\ref{toym}), we make the minimal choice of a quartic (in the field $\phi$) 
potential, with quartic coupling $u_k$, and a degenerate minimum  at  $\rho=\rho_0$,
\begin{equation}\label{poten}
 V_k (\phi )   =  \frac{u_k}{8} \, \left(  |\phi |^2  - \rho_0 \right)^2 \,\,.
 \end{equation}

By representing the field in polar coordinates
\begin{equation}\label{polarcoo}
 \phi(\br) = \sqrt{\rho(\br)} \,\, e^{i\vartheta(\br)}
\end{equation}
and rewriting  $\rho$ in terms of the field expectation value plus a fluctuation term, $\rho(\br) = \rho_0 + \tilde\rho(\br)$, we can rearrange Eq.\,(\ref{toym}) 
 in a simpler form. 
In fact, if we look at the infrared regime below the scale set by the expectation value $\rho_0$,  we can neglect the effect of the fluctuations $\tilde\rho (\br)$ that are 
suppressed with respect to $\rho_0$, while we retain the fluctuations of the angular field $\vartheta (\br)$, and we get the following effective action in the infrared regime
\begin{eqnarray} \label{toyir}
&&\Gamma^{IR}_k[\phi] =\int d^4\br  \Bigg\{   \left (a_1\rho_0+a_2 \,\rho_0^2 \right ) \, \left ( \partial \vartheta \, \partial \vartheta  \right )
\nonumber\\
&+&\left (b_1\rho_0+b_2 \,\rho_0^2 \right) \,  \left ( \partial^2 \vartheta \,\,\partial^2 \vartheta \right ) \nonumber \\
&+ &\left ( \frac{b_1}{\rho_0}  +b_2+4b_3-2b_4+b_5+b_6+b_7 \right ) \rho^2_0 \left( \partial \vartheta \,\partial \vartheta \right)^2  \Bigg\}
\end{eqnarray}

We observe that neither the potential, nor the term proportional to $a_3$
appear in  Eq.\,(\ref{toyir}), as they yield contributions 
proportional to $\tilde\rho (\br)$ or  $\partial \tilde\rho (\br)$, that are discarded. Instead, terms with two derivatives in Eq.\,(\ref{toym}) 
generate in  Eq.\,(\ref{toyir}) quadratic terms in $\vartheta$, and those with four derivatives yield both quadratic and quartic terms in $\vartheta$.

If the terms proportional to  $\left( \partial \vartheta \,\partial \vartheta \right)^2 $ in  Eq.\,(\ref{toyir}) were absent, the remaining effective
action would be quadratic in the angular field $\vartheta$, leading to a simple infrared behavior of the theory,
which is  instead the case of the truncation of the effective action studied in \cite{jame}, where only two derivatives of $\phi$ are retained, 
i.e.  in Eq.\,(\ref{toyir}) all $b_i=0$. In fact, in this case  the infrared effective  action is quadratic in $\vartheta$, as all the parameters $a_i$ 
are related to the operator $\left( \partial \vartheta \,\partial \vartheta \right)$.
Therefore, the general truncation in Eq.\,(\ref{toym}) does not produce an infrared effective action quadratic in $\vartheta$,
and even the more restricted approximation involving only the two operators proportional to $W_A$ and $W_B$ generates
terms that are quartic in $\vartheta$. 

However, from Eq.\,(\ref{toym}) it is also evident that there are specific combinations of the parameters $b_i$ that cancel the coefficient of the operator  
$\left( \partial \vartheta \,\partial \vartheta \right)^2$, yet with  a non-vanishing coefficient of $\left( \partial^2 \vartheta \,\partial^2 \vartheta \right)$.
If this fine tuning is realized, then the corresponding effective theory in the infrared is a quadratic theory in the angular field $\vartheta$,
with a coordinate independent background $\rho_0$ and, under these conditions, it is straightforward to determine the algebraic decay of 
the correlator of the field $\phi$ in the same way as for the BKT phase \cite{jame,popov87}.

In fact, if we look at the large distance (or infrared) behavior of the correlation function 
\begin{equation}\label{correl}
\left \bra \phi(\br) \phi^*(\b0) \right \ket =
\left  \bra \sqrt{\rho(\br) \rho(\b0)}\;\, e^{ i \, [\vartheta(\br) - \vartheta(\b0)]} \right \ket \, ,
\end{equation}
we are allowed to compute the average by making use of the effective action in Eq.\,(\ref{toyir}) with both $\rho(\br)$ and  $\rho(0)$  replaced by $\rho_0$,
as the fluctuations of the modulus around $\rho_0$ are neglected.  In addition, if the  quartic term in  $\vartheta$ is cancelled by a suitable choice of the $b_i$ 
as discussed above, the action is quadratic in the angular field, and Eq.\,(\ref{correl}) becomes
\begin{equation}\label{correl2} 
\left \bra \phi(\br) \phi^*(\b0) \right \ket =
\rho_0 \; {\rm exp}\left [ {-\frac{1}{2} \left \bra \left [ \vartheta(\br) - \vartheta(\b0) \right ]^2 \right \ket}  \right ] \, ,
\end{equation}
where the expectation value of the squared angular field is:
\begin{equation}
\label{propa}
\left \bra [ \vartheta(\br) - \vartheta(\b0) ]^2 \right\ket =
\int \frac{d^4\bq}{(2\pi)^4} \;\frac{ |e^{i\bq\br} - 1|^2 }{ \left[ 2 \rho_0\,(b_1\,q^4+ a_1\,q^2 )\right ]  }
\end{equation}
and the factor $2 \rho_0$ in the denominator in Eq. (\ref{propa}) comes from the normalization
of the terms quadratic in $ \vartheta$  in Eq. (\ref{toyir}).

To avoid lengthy expressions, in Eq.\,(\ref{propa}) we chose, without loss of generality,
$a_2=b_2=0$  in  the $\vartheta$ propagator, and after performing the angular integration we get ($r\equiv |\br |$)
\begin{equation} \label{propared}
 \left \bra [ \vartheta(\br) - \vartheta(\b0) ]^2 \right\ket =
 \frac{2}{\eta_1}
  \int_0^{\Lambda r}  {\rm d}x    \;\frac{   x^3 \left[   1- (2/x) \, J_1(x) \right ] }{ \left[  x^4+ (a_1 r^2/ b_1) \,x^2 )\right ]  }
\end{equation}
where $\Lambda$ is an ultraviolet cutoff on the momentum,  $J_1(x)$ is the Bessel function of the first kind of order 1,
and we defined 
\begin{equation} \label{eta1}
\eta_1=\frac{1}{16\,\pi^2\,\rho_0\,b_1} \,.
\end{equation}
Due to the asymptotic form of the Bessel function at large $x$: $J_1(x) \sim \sqrt{2/(\pi\,x)}\,{\rm cos}(x-3\pi/4)$ and at small $x$: $J_1(x) \sim   x/2  -x^3/16$, one easily realizes that the part of the 
integrand proportional to $J_1(x)$ is regular and produces a finite contribution to the integral, while the remaining part of the integrand vanishes at $x=0$, but 
yields a logarithmic divergence at large $x$, made finite by the insertion of $\Lambda$,  so that, after integrating, the dominant term is $(2/\eta_1) \ln(\Lam  r) $.

We notice that the parameter $a_1$, associated to the operator containing  two derivatives of $\vartheta$, does not contribute to the leading result, while the addition of the term proportional to $b_2$,
which was discarded in  Eq.\,(\ref{propa}), would have simply produced a redefinition of $\eta_1$.
When the leading term of Eq.\,(\ref{propared}) is inserted in Eq.\,(\ref{correl2}), we recover the algebraic large distance decay of order parameter correlation  
\begin{equation}
\left \bra \phi(\br) \phi^*(\b0) \right \ket 
 \propto \frac{\rho_0}{r^{\,\eta_1}} \quad 
\end{equation}
with the exponent $\eta_1$ inversely  proportional to the product of the renormalization factor of the four derivative operator,  $b_1$, times the symmetry breaking scale, $\rho_0$.
This finding  is essentially the same of the result obtained for  the BKT phase \cite{popov87,jame}, with  $b_1$  replaced by the renormalization factor of the two-derivative
term $\left ( \partial \vartheta \, \partial \vartheta  \right )$.

%%%%%%%%%%%%%%%%%%%%%%%%%%%%
\vskip 8 pt
{\section{Flow equations}
\label{III}}

Now we  turn to the standard cartesian representation of the complex field $\phi(\br)$ by decomposing it into a longitudinal and a transverse component and, 
as we are interested to explore the phase of the theory that could possibly display properties similar to the low temperature BKT phase, 
we  separate in the longitudinal component  the constant term corresponding to the  minimum of the potential in  Eq.\,(\ref{poten}), $\alpha=\sqrt{\rho_0}$:  
\begin{equation}
 \label{cartesian}
 \phi(\br) = \alf + \sg(\br) + i\pi(\br) \; .
\end{equation}

Then, instead of considering the general structure of the effective action in Eq.\,(\ref{toym}), it is convenient to start from the simple ansatz considered in \cite{jame}
for the BKT phase, suitably rearranged to the LP case, so that the particular truncation here adopted, reads :
\begin{eqnarray} \label{toymbis}
&\Gamma_k [\phi] &=\int d^4\br  \Bigg\{ \frac{u_k}{8} \, \left( |\phi |^2 - \alpha^2_k \right)^2 +
 \frac{W^A_k}{2} \left [ \partial^2 \phi \, \partial^2  \phi^* \right ] 
\nonumber \\ 
&&+ \frac{W^B_k}{8} \,  \left [ \partial^2 |\phi |^2 \right]^2  + \frac{Z^A_k}{2} \left [ \partial \phi \, \partial  \phi^* \right ] +  \frac{Z^B_k}{8}  \,  \left [ \partial |\phi |^2 \right]^2
  \Bigg\}
\end{eqnarray}
where we used the potential (\ref{poten}) and kept, together with the  $O(\partial^4)$,  also the $O(\partial^2)$ operators, both with the same kind of 
parametrization.  Although Eq.\,(\ref{toymbis})  does not correspond to a complete parametrization to order $O(\partial^4)$ and to the fourth power of $\phi$,
which is instead given in  Eq.\,(\ref{toym}),  yet it includes all the operators in Eq.\,(\ref{toym}) that contribute to the propagator of the two real fields $\sg(\br)$ and $\pi(\br)$. 
In fact, once $\phi$ is replaced by the expression in Eq.\,(\ref{cartesian}) that contains the coordinate independent term $\alpha_k$, those operators not included in  Eq.\,(\ref{toymbis})
do appear only in three or four point functions and not in the two point functions, due to the effect of the derivatives of the field $\phi$. 
Our goal is to check whether this truncation still maintains those features of the BKT phase that are illustrated in \cite{jame}.

The flow of the effective action  is determined by the FRG equation and therefore the various parameters in Eq.\,(\ref{toymbis}),  namely 
(in the following, for simplicity we omit the subscript $k$ of the various parameters) $u$, $\alpha$, $W^A$, $W^B$, $Z^A$, $Z^B$,  
are obtained by solving Eq.\,(\ref{rgfloweq}) for specific initial conditions given at a  large value of the scale $k=K_{UV}$.
Clearly, one has to select a particular procedure to extract  from Eq.\,(\ref{rgfloweq}) the flow equation of each parameter and, in the case of the quartic coupling,
it turns out to be more convenient to replace  the flow equation of $u$ with that of  the $\sg$ field mass, defined as the coefficient of $\sg^2$ in the effective potential of our model,
that is
\begin{equation} \label{mualf}
 m_{\sg}^2 = u \alf^2 \; ,
\end{equation}
while for the coefficients of the  $O(\partial^4)$  and of  the $O(\partial^2)$ operators, the rearrangement
\begin{eqnarray} \label{w&z}
 &&W_{\sg} = W^A+ W^B \alf^2 \;\;\; \; ; \;\;\;\;\; Z_{\sg} = Z^A + Z^B \alf^2  \nonumber \\
&& W_{\pi} = W^A \;\;\; \;\;\;\;\;\;\;\;\;\;\;\; \;\; ; \;\;\;\;\;\; Z_{\pi} = Z^A
 \end{eqnarray}
yields the following simple form of the two point functions of the fields $\sg$ and $\pi$ (by construction the field $\pi$ is massless):
\begin{eqnarray}
\label{gammas}
 \Gamma^{(2)}_\sg(q) &=& W_{\sg}\, q^4 + Z_{\sg}\, q^2 + m_\sg^2  \, , \\
\label{gammap}
 \Gamma^{(2)}_\pi(q) &=& 
W_{\pi}\, q^4 + Z_{\pi}\, q^2  \, .
\end{eqnarray}

To derive the FRG equations of $m_{\sg}$, $\alpha$, $W_{\pi }$, $W_{\sg}$, $Z_{\pi}$, $Z_{\sg}$, we follow the approach displayed in \cite{jame} (see also \cite{metz})
in which  the flow of these parameters is extracted from the FRG equations of the two point functions $ \Gamma^{(2)}_\sg(q)$ and  $\Gamma^{(2)}_\pi(q)$ which, in turn,  come from 
Eq.\,(\ref{rgfloweq}), with the exception of $\alpha$, whose equation follows from the condition that the effective action must have a minimum at $|\phi |^2=\alpha^2$ at each 
value of the scale $k$  (in the following we  use the notation  $\int_\bq \equiv \int \frac{d\bq^4}{(2\pi)^4}$) :
\begin{equation} \label{alfflow}
 \frac{d\alf^2}{dt} = - \,  \int_\bq   \,
  \left\{ \left [1 + \frac{2U(q)}{u} \right ] \, G'_\sg(q) +  G'_\pi(q) \right \} \, 
\end{equation}
where we made use of the relation (\ref{mualf}), introduced the momentum-dependent vertex
\begin{equation} 
\label{Uvertex}
U(q) = u + W^B \,q^4 +Z^B\, q^2 \; ,
\end{equation}
and we also introduced the propagators of $\sg$ and $\pi$ modified by the infrared regulator $R(q)$
(here the subscript $\tau$ stands either for $\sg$ or for $\pi$) :
\begin{equation} 
\label{propsp}
 G_{\tau}(q) = \left [  \Gamma^{(2)}_{\tau}(q) + R(q) \right ]^{-1}  \; .
\end{equation}
Finally, we adopt the notation used in \cite{jame} of indicating with a prime 
the derivative of the regulator $R$ with respect to $t$; i.e. for a generic function $F(R(q))$ one has
\begin{equation} \label{DR}
 F'(R(q))=  \frac{dF(R)}{dR} \,\frac{dR}{d t }\, .
\end{equation}

The structure of the flow equation of the the two point functions $ \Gamma^{(2)}_\sg(q)$ and  $\Gamma^{(2)}_\pi(q)$ with our effective action  is exactly the same as 
the one obtained for the two point functions of the BKT phase, derived in Ref. \, \cite{jame}. 
Therefore in our case, after the required changes, we get for  the longitudinal component $ \Gamma^{(2)}_\sg(p)$ 

\begin{eqnarray} \label{gamma_s}
&& \frac{d \Gamma^{(2)}_\sg(p)}{dt } = 
 \frac{1}{2} \left[ u + 2U(p) \right]  \frac{d\alf^2}{dt}
 \nonumber \\
 &+& \frac{1}{2} \int_\bq 
 \Bigl\{ \left[ u + 2U(p+q) \right] \, G'_{\sg}(q) 
 + u \, G'_{\pi}(q) \Bigr\}
 \nonumber \\
 &-& \frac{\alf^2}{2} \int_\bq 
 \left[ U(p) + U(q) + U(p+q) \right]^2  \left[ G_{\sg}(q) G_{\sg}(p+q) \right]'
  \nonumber \\
 &-& \frac{\alf^2 }{2} \int_\bq 
 \left[ U(p) \right]^2 \,
\left[ G_{\pi}(q) G_{\pi}(p+q) \right]' \,  ,
\end{eqnarray}
and for the transverse component $\Gamma^{(2)}_\pi(p)$:
\begin{eqnarray} \label{gamma_p}
&& \frac{d\Gamma^{(2)}_\pi(p)   }{dt}  = 
 \frac{u}{2}  \frac{d\alf^2}{dt} 
 \nonumber \\
 &+& \frac{u}{2} \int_\bq
 \left\{ \left[ 1 + \frac{2U(p+q)}{u} \right] \, G'_{\pi}(q) 
 + G'_{\sg}(q) \right\} 
 \nonumber \\
 &-& \alf^2  \int_\bq   \left[ U(q) \right]^2 \,
 \left[ G_{\sg}(q) G_{\pi}(p+q) \right] ' \, .
\end{eqnarray}

It must be noticed that the first terms in the right hand side of both Eqs. \,(\ref{gamma_s}) and  \,(\ref{gamma_p}),
come from the dependence of the two point function on the parameter $\alf^2$, which in turn carries a dependence on the running scale $t$
and therefore contribute to the flow equation.

From Eqs. \,(\ref{gamma_s}),  \,(\ref{gamma_p}), one directly derives the flow of the other parameters. In fact, the flow of the squared mass $m_\sg^2 $
is simply obtained by taking the momentum independent projection of the equation for the longitudinal two point function, i.e. by putting $p=0$ in Eq. \,(\ref{gamma_s}):
\begin{eqnarray} \label{m_s}
 \frac{d m_{\sg}^2 }{dt} &=& 
 \frac{3u}{2} \frac{d\alf^2 }{dt} +
 \frac{1}{2} \int_\bq 
 \Bigl\{ \left[ u + 2U(q) \right] \, G'_{\sg}(q) 
 + u \, G'_{\pi}(q) \Bigr\} \quad
 \nonumber \\
 &-& \frac{\alf^2 }{2} \int_\bq 
 \left\{ \left[ u + 2U(q) \right]^2 G_{\sg}^2(q)
 + u^2 \, G_{\pi}^2(q) \right\}'   \,.
\end{eqnarray}

The flow equation of the parameters $W$ and  $Z$, due to their definition in Eqs. \,(\ref{gammas}),  \,(\ref{gammap}), are  obtained by selecting the 
coefficients of $p^4$  (for $W$) and of $p^2$ (for $Z$) in  Eqs. \,(\ref{gamma_s}),  \,(\ref{gamma_p}). This requires an expansion
in powers of the external momentum $p$, of the various terms appearing in the right hand side of Eqs. \,(\ref{gamma_s}),  \,(\ref{gamma_p}).
Therefore, by indicating with a subscript $p^4$ or $p^2$ the operation of selecting only the coefficient of that particular power of $p$,
the flow of $W_\sg$ reads:
\begin{eqnarray} \label{W_s}
&& \frac{d  W_{\sg} }{dt} =
W^B  \; \frac{d\alf^2}{dt} + W^B \int_\bq G'_{\sg}(q)
 \nonumber \\
 &-& \frac{\alf^2}{2} \Biggl [ \int_\bq
 \left[ U(p) + U(q) + U(p+q) \right]^2  \left[ G_{\sg}(q) G_{\sg}(p+q) \right] '  \Biggr ]_{p^4}
   \nonumber \\
&-& \frac{\alf^2}{2} \Biggl [   \int_\bq \left[ U(p) \right]^2 
 \left[ G_{\pi}(q) G_{\pi}(p+q) \right]'  \Biggr ]_{p^4}
\end{eqnarray}
while the flow for $Z_\sg$ is:
\begin{eqnarray} \label{Z_s}
&& \frac{d  Z_{\sg} }{dt} =
Z^B \;  \frac{d\alf^2}{dt} +\int_\bq ( Z^B +  3\,W^B \, q^2 )\, G'_{\sg}(q)
\nonumber \\
 &-& \frac{\alf^2}{2} \Biggl [ \int_\bq
 \left[ U(p) + U(q) + U(p+q) \right]^2  \left[ G_{\sg}(q) G_{\sg}(p+q) \right] '  \Biggr ]_{p^2}
   \nonumber \\
&-& \frac{\alf^2}{2} \Biggl [   \int_\bq \left[ U(p) \right]^2 
 \left[ G_{\pi}(q) G_{\pi}(p+q) \right]'  \Biggr ]_{p^2}
\end{eqnarray}
where $W^B$ and $Z^B$ are expressed in terms of $W_\sg$, $W_\pi$,  and  $Z_\sg$, $Z_\pi$ through the relations in \,(\ref{w&z}).

The flow equation for $W_\pi$, obtained from Eq.\,(\ref{gamma_p}), can be further simplified, \cite{jame},
by using the relation $\alf^2 U(q) = G_\sg^{-1}(q) - G_\pi^{-1}(q)$, thus obtaining:
\begin{equation} \label{W_p}
\frac{dW_\pi}{dt}  =  \frac{1}{\alf^2} 
\Biggl [ \int_\bq   \Big[ G_\pi^{-1}(q) G_\pi(p+q)
 - G_\sg(q) G_\pi^{-2}(q)  G_\pi(p+q) \Big]'   \Biggr ]_{p^4} 
\end{equation}
and, similarly, the flow equation for $Z_\pi$ reads:
\begin{equation} \label{Z_p}
 \frac{dZ_\pi}{dt}  = \frac{1}{\alf^2} 
  \Biggl [ \int_\bq   \Big[ G_\pi^{-1}(q) G_\pi(p+q)
 - G_\sg(q) G_\pi^{-2}(q)  G_\pi(p+q) \Big]'   \Biggr ]_{p^2} 
\end{equation}
\vskip 8 pt

The set of six flow equations (\ref{alfflow}),\,(\ref{m_s} -
\ref{Z_p}), represents the output of the
specific truncation made on the effective action in Eq.\,(\ref{toymbis}).
It is a closed set, as it is possible to get rid of the coupling $u$ through Eq.\,(\ref{mualf}),
as well as of the field renormalization parameters $W^B$ and $Z^B$ with the help of 
Eqs.\,(\ref{w&z}), without explicitly computing the flow of these quantities.
The numerical analysis of the six flow equations is presented in Sec.\,\ref{Numerical}.

%%%%%%%%%%%%%%%%%%%%%%%%%%%%%%%%%%%%%%%%%%%%%%%%%%%%

\vskip 8 pt
{\section{Transverse component approximation}
\label{IV}}

Before turning to the numerical analysis,  it is instructive to observe that  a fixed point solution occurs in a simpler framework.  This further simplification consists  in neglecting the  effects of the  longitudinal 
fluctuations, i.e. in neglecting the flow of  $W_\sg$, $Z_\sg$,  $m^2_\sg$ and  discarding $G_\sg(q)$ from the remaining equations. Then,  by retaining only the transverse fluctuations, one can check that the stiffness $J$, 
which is defined  as the product of the  transverse field renormalization constant  (for the LP, it is given by  $W_\pi$) times the field  expectation value
\begin{equation} \label{stiffness}
J=W_\pi\,\alf^2 \, ,
\end{equation}
does not depend on $t$ and therefore, each different value  of $J$ corresponds to a fixed point solution.

In order to illustrate this point we need to search for $t$-independent solutions of the flow equations of the parameters, properly rescaled in units of the 
running scale $k= K_{UV}\, e^{-t}$,  that, for the scaling close to a LP means:  ${\rm a}^2=k^{-\eta}\alf^2$, \, $w_\pi =k^{\eta}\, W_\pi$\,, \; $z_\pi =k^{\eta-2}\, Z_\pi$.
The corresponding FRG equations are (the momenta are also rescaled in units of $k$ so that the variable $\widetilde q$ in the following integrals is dimensionless)
\begin{equation} \label{fp_eq1}
 \partial_t {\rm a}^2 -\eta {\rm a}^2 = (\eta-4) \, w_\pi \,\int_{\widetilde \bq} \widetilde G_\pi^{\,2}
\end{equation}
\begin{equation} \label{fp_eq2}
 \partial_t  w_\pi + \eta w_\pi = \frac{(4-\eta )\,  w^2_\pi}{ {\rm a}^2 }\int_{\widetilde \bq} \widetilde G_\pi^{\,2}
\end{equation}
\begin{equation} \label{fp_eq3}
 \partial_t  z_\pi + (\eta-2)\,  z_\pi =  \frac{(4-\eta )\,  w_\pi}{ {\rm a}^2 }   \int_{\widetilde \bq} (z_\pi +3 w_\pi {\widetilde q}^{\,2} ) \,\widetilde G_\pi^{\,2}                        
\end{equation}
and the rescaled propagator is
\begin{equation} \label{fp_pr}
\widetilde G_\pi=\left [ w_\pi \, ( {\widetilde q}^{\,4} +1) + z_\pi \, {\widetilde q}^{\,2}  \right ]^{-1}
 \end{equation}
where the term not depending on ${\widetilde q}$ comes from the rescaling of the regulator $R$, for which we made the minimal choice (but sufficient for our purpose) :
$ R = W_\pi \, k^4 $.

Stationary (i.e. $t$-independent) solutions of this set of three equations provide  the fixed points of the simplified problem. Moreover, to get rid of the 
redundant overall multiplicative factor in  the effective action we must set $w_\pi=1$, and we are left with  the three independent parameters $\eta, \, z_\pi$ and ${\rm a}^2$.
However, it is easy to realize that Eqs.\,(\ref{fp_eq1}) and (\ref{fp_eq2}) are not independent, and therefore one of the three parameters  cannot be constrained. 
Since, according to the definition in Eq.\,(\ref{stiffness}), the stiffness is $J={\rm a}^2$, it is convenient to solve  Eqs.\,(\ref{fp_eq2}), (\ref{fp_eq3}) 
in terms of the free parameter ${\rm a}^2$, thus obtaining the two fixed point conditions
\begin{eqnarray} \label{fp1}
&& \eta\,{\rm a}^2 = (4-\eta) \, \,\int_{\widetilde \bq}  \,\left [ {\widetilde q}^{\,4} + z_\pi \, {\widetilde q}^{\,2}  +1 \right ]^{-2}
\\ \label{fp2}
&& z_\pi\,{\rm a}^2 = (\eta-4)\,\int_{\widetilde \bq} \,\, \frac{3\,{\widetilde q}^{\,2} }{2} \,\left [ {\widetilde q}^{\,4} + z_\pi \, {\widetilde q}^{\,2}  +1 \right ]^{-2} 
\end{eqnarray}
which represent a continuous line of fixed point solutions for both $\eta$ and $z_\pi$,  parametrized by the stiffness $J$, exactly as it happens for the BKT phase, \cite{jame}. 

\begin{figure}[tb]
\begin{center}
\includegraphics[width=8cm]{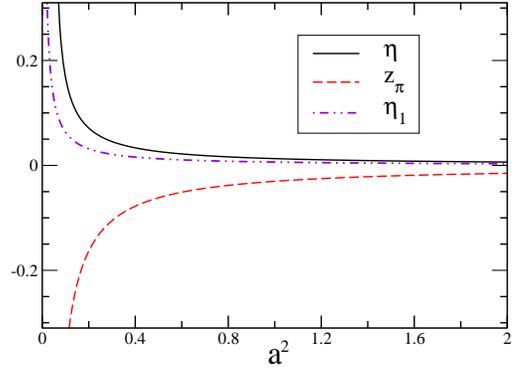}
\caption{Fixed point values of $\eta$, $z_\pi$ and $\eta_1$ (see Eqs.\,(\ref{fp1}), (\ref{fp2}), (\ref{eta1})) as functions of the stiffness. }
\label{figr1}\end{center}
\end{figure}

This line of solutions is displayed in Fig. \ref{figr1}, and one immediately observes that while $\eta$ is positive, $z_\pi$ is negative, with  both parameters vanishing for large values of ${\rm a}^2$ and both 
showing a divergent behavior at small ${\rm a}^2$.  Therefore, the fixed point solution requires a negative value of  $z_\pi$ to compensate the effect of the ${\widetilde q}^{\,4}$  term in the propagator.
For comparison, in Fig. \ref{figr1} we  also plotted the anomalous dimension shown in Eq.\,(\ref{eta1}) by identifying ${\rm a}^2=J=   2 b_1 \rho_0$,  in accordance with Eqs.
(\ref{toyir}), (\ref{toymbis}), (\ref{stiffness}), and qualitative agreement with $\eta$ coming from Eqs.\,(\ref{fp1}), (\ref{fp2})  is observed. (Incidentally, by setting  ${\rm a}^2= 2 b_1 \rho_0$ and 
$z_\pi = 0$, the integral in in Eq. (\ref{fp1}) is performed analytically and we recover again the relation given in Eq. (\ref{eta1}) ).

Concerning the extension of the range of ${\rm a}^2$ for which there are fixed point solutions, one finds that  there is no limit for ${\rm a}^2\to \infty$ 
and the corresponding solutions $\eta,\,z_\pi \to 0$,
while it is easy to understand that  a singularity in the propagator does show up  at small ${\rm a}^2$, due to the large negative value of $z_\pi$, thus 
interrupting the  line of solution at about ${\rm a}^2=0.065$ with $\eta=0.402$ and      $z_\pi=-0.824$ . However we do not expect that this endpoint could 
be related to a possible phase transition to the disordered phase because  it  is originated by a singularity in  Eqs.\,(\ref{fp1}), (\ref{fp2}) 
which  are the result of an approximation that becomes less reliable at small ${\rm a}^2$. 

In fact, according to the  results obtained in $d=2$, this approximation  does not predict a  critical point of the stiffness $J$ that indicates the transition  
from the BKT phase to the disordered phase,  despite it captures many peculiar properties of the BKT phase at large $J$.  
Therefore, one could expect that also in $d=4$, a similar transition could  possibly occur at some value of  ${\rm a}^2$, larger than the one at which the coupled equations (\ref{fp1}), (\ref{fp2})  
no longer have a fixed point solution.

Now we turn  to the study of the eigendirections of equations \,(\ref{fp_eq1}), \,(\ref{fp_eq2}), (\ref{fp_eq3}), around the fixed point solutions.
This is realized by considering  the following small corrections to the  stationary solutions, which present a dependence on the running parameter $t$
through  an exponential  factor:  
$\delta {\rm a}^2\exp[\lambda t ]$,\,
$\delta w_\pi \exp[\lambda t ]$,\,
$\delta z_\pi \exp[\lambda t ]$,  and then by solving the set of three equations, linear in  the perturbations $\delta {\rm a}^2$,\, $\delta w_\pi$,\, $\delta z_\pi$, 
that are obtained from  an expansion of Eqs. \,(\ref{fp_eq1}), \,(\ref{fp_eq2}), (\ref{fp_eq3})  around the fixed point solution.  
This procedure amounts to the resolution of the eigenvalue problem for the set of three perturbations with  eigenvalue $\lambda$, whose sign, positive or negative (or vanishing), 
characterizes  the associated  eigenstate as relevant or irrelevant (or marginal). In fact, when $t\to \infty$,  a positive $\lambda$ corresponds to a growing, and therefore relevant,  
perturbation, while  negative (zero) values of  $\lambda$ give decreasing (constant) i.e. irrelevant (marginal) perturbations.

This analysis turns out to be rather simple if we focus on the region of very large ${\rm a}^2$ where, according
to Eqs.\,(\ref{fp1}) and  (\ref{fp2}), the fixed point solutions for $\eta$ and $z_\pi$ are of order $1/{\rm a}^2$ and we can treat these three quantities as  $O(\epsilon)$ terms, with $\epsilon\equiv 1/{\rm a}^2$.
Then, the resolution of the linear equations to $O(1)$ order, gives one relevant solution with $\lambda=2$, and two marginal solutions with $\lambda=0$.
If the $O(\epsilon)$ effects are included, the three eigenvalues become
\begin{eqnarray} \label{eigen}
&&\lambda_1=2-\frac{24}{ {\rm a}^2} \,\int_{\widetilde \bq} \,\, {\widetilde q}^{\,4}  \,\left [ {\widetilde q}^{\,4} + z_\pi \, {\widetilde q}^{\,2}  +1 \right ]^{-3} 
\nonumber\\
&&\lambda_2=\eta  
\nonumber\\
&&\lambda_3=-\eta
\end{eqnarray} %
which are respectively very close to the scaling dimensions of $Z_\pi$, $\alpha^2$, and $ W_\pi$.

Before concluding this section, two comments are in order. First, we observe that the  fixed point solutions here shown are strictly related to the $O(2)$  (in replacement of the original $U(1)$) symmetry
of the model. In fact, the presence of additional transverse components of a $O(N)$ theory would produce a multiplicative factor $(N-1)$ in the right hand side of Eq.\,(\ref{fp_eq1}),
but leaving unchanged Eqs.\,(\ref{fp_eq2}) and  (\ref{fp_eq3}), because the factor $(N-1)$ would only replace the term $< 1 >$ in the square brackets in the second line of Eq.\,(\ref{gamma_p}),
and this term  does not contribute to the flow equations of $Z_\pi$ and $ W_\pi$ i.e. to Eqs.\,(\ref{fp_eq2}), (\ref{fp_eq3}). 
Therefore, additional transverse components would spoil the  relation between Eqs.\,(\ref{fp_eq1}) and  (\ref{fp_eq2}), that is essential in determining the fixed point solutions.

The second comment concerns  the appearance of a relevant direction related to the parameter $Z_\pi$, with eigenvalue $\lambda_1 \sim 2$
that is  peculiar of this line of fixed points and has no counterpart in the  analysis of  2-dimensional BKT phase.

 %%%%%%%%%%%%%%%%%%%%%%%%%%%%%%%%%%%%%%%%%%%%%%%%%%%%%%%%%%%%%%%%%%%%
\vskip 8 pt

{\section{TRANSVERSE AND LONGITUDINAL COMPONENTS}
\label{Numerical}}

After the determination of a line of fixed point for the simplified set of flow equations, we include the effects of the longitudinal fluctuations and analyze the 
full set,  Eqs.\, (\ref{alfflow}),\,(\ref{m_s} - \ref{Z_p}). One can easily check that now $\alpha^2$ and $W_\pi$ do not show a compensating behavior such that their product,
$J= W_\pi\, \alpha^2$, is $t$-independent, as observed when the longitudinal fluctuations are neglected. Therefore, at least in principle, the line of fixed point is no longer present.

\begin{figure}[tb]
\begin{center}
\includegraphics[width=8cm]{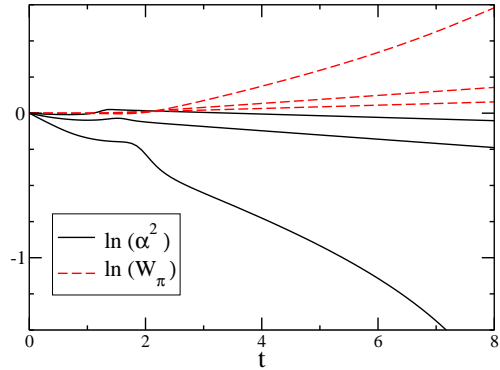}
\caption{Flow of ${\rm ln}(\alf^2)$  (solid, black online) and $ {\rm ln}( W_\pi)$  (dashed, red online), with initial values  $\underline \alf^2=$ 1.1 (the two inner curves), 
0.5 (two intermediate curves), 0.24 (two external curves).}
\label{figr2}\end{center}
\end{figure}

Then,  from the analysis \cite{jame}, it is known that the inclusion of the longitudinal fluctuations produces the following effect for the two-dimensional case:  the line of fixed points,
peculiar of the BKT phase, is only partially observed in this approximation, and the flow of $J$ is  practically stationary only at very large initial values of $J$ (which corresponds to the small temperature
regime of  \cite{jame}), but when the flow starts at smaller $J$,  this parameter, after remaining almost constant in a long range of $t$,  decreases rapidly 
at some, still large, value of  $t$.
When starting the flow at even smaller $J$, the almost constant region shrinks and tends to disappear, so that not even an approximate fixed point is observed. 
Consequently, it is natural to expect that in the four-dimensional problem, the same truncation could at most lead to a similar picture with an approximated line of fixed points.

Therefore, we solve numerically the flow equations  (\ref{alfflow}),\,(\ref{m_s} - \ref{Z_p}) for the parameters  $\alf^2$, $m_\sg^2$, $W_\pi$,  $W_\sg$, $Z_\pi$  and $Z_\sg$,
with the 'time' $t={\rm ln}(K_{UV}/k)$ running from $t=0$ to large positive values (infrared region)  and, in particular, we monitor the $t$-dependence of the dimensionless stiffness $J$,
as the approaching  of a fixed point requires an asymptotically constant $J$.
The numerical analysis is more easily  performed by using the minimal  infrared regulator introduced before, $ R = W_\pi \, k^4 $, which does not depend on the propagator momentum.

The flow equations are studied for different  initial values of the six parameters, and in all cases the two field renormalizations that set the overall normalization of the action, are taken 
$\underline W_\pi=\underline W_\sg=1$ at the starting point of the flow, $t=0$ (in the following, underlined  quantities  indicate the particular value taken by such parameters at $t=0$). 
Consequently  the spanning of $J$ is obtained by taking different initial values $\underline\alf^2$.

The choice of the initial value of the remaining parameters requires some care. In fact, once $\underline \alf^2$ is fixed,  the flow typically stops at finite $t$, because some  parameter goes to zero  or to infinity, unless
$\underline m_\sg^2$, $\underline Z_\pi$  and $\underline Z_\sg$  are suitably tuned: only for a specific choice of these parameters the running scale $t$ can grow to much  larger values. This effect is not surprising 
as it is known that the flow reaches  a fixed point  only if all the relevant parameters  are taken on the critical manifold; otherwise  the flow is driven away from the fixed point in the infrared limit,
and in our particular case,   $m_\sg^2$, $Z_\pi$  and $Z_\sg$  are expected to be relevant parameters from an inspection of their scaling dimensions.

\begin{figure}[tb]
\begin{center}
\includegraphics[width=8cm]{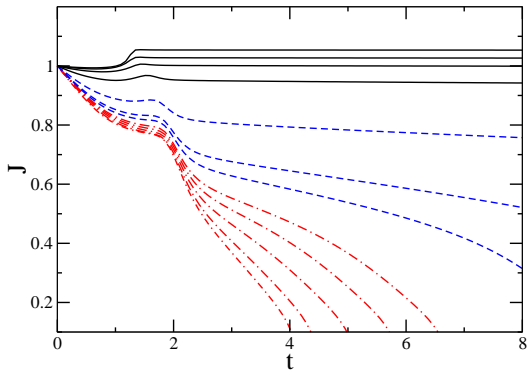}
\caption{Flow of the stiffness $J$ for values of 
$\underline \alf^2$ (from top to bottom) :  1.5, 1.1, 0.8, 0.5 (solid, black online, curves),  
0.3, 0.25, 0.24 (dashed, blue online),  0.23, 0.226, 0.22, 0.21, 0.2 (dot-dashed, red online).}
\label{figr3}
\end{center}
\end{figure}

In Fig. \ref{figr2} the evolution with $t$ of the logarithm of the two parameters $\alf^2$, and $W_\pi$, both normalized to their respective initial value, is reported for three different initial values.
In particular the two almost flat lines, solid (black online) for ${\rm ln}(\alf^2)$ and dashed (red online) for $ {\rm ln}( W_\pi)$, are obtained for  $\underline \alf^2=1.1$ (all dimensionful quantities are expressed in 
units of the energy scale $K_{UV}$, corresponding to  $t=0$), while the contiguous steeper couple of lines correspond to $\underline\alf^2=0.5$ and the remaining steepest couple of lines to $\underline \alf^2=0.24$. 
We observe that, after a transient regime, the first two couples of curves become practically  straight with opposite slope, which indicates a constant stiffness $J$, 
but at smaller $\underline \alf^2$ the curves show some deviation from a straight line, at large $t$.

A more detailed picture of  the stiffness $J$, that confirms the indications of Fig. \ref{figr2},  is reported  in  Fig. \ref{figr3} for twelve different values of $\underline \alf^2$, that can be arranged in three groups. 
Namely, going from top to bottom, the first group is  (1.5, 1.1, 0.8, 0.5) corresponding to solid (black online) curves, the second is (0.3, 0.25, 0.24) with dashed (blue online) curves,  and finally 
the third  is (0.23, 0.226, 0.22, 0.21, 0.2) with dot-dashed (red online) curves.
For the first group we observe straight lines (up to $t=8$),  with the lowest case showing a very slight  slope, while in the second group the slope is more pronounced and the curves start to be turned downwards at 
large $t$.  The last group correspond to curves that are no longer straight and reach zero at  $t < 8$.
Therefore, for the first group a quasi-fixed point regime is observed, that is eventually altered at some extremely low energy scale, whereas  the  quasi-fixed point features  are totally absent for the last group of curves.
The second group has an intermediate behavior and it is evident that $J$ reaches zero at some, not too far, value of $t$.

\begin{figure}[tb]
\begin{center}
\includegraphics[width=8cm]{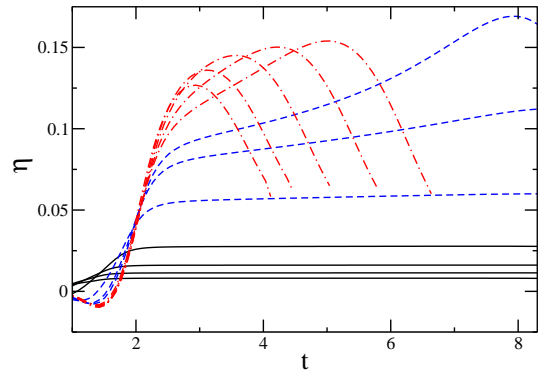}
\caption{Anomalous dimension $\eta$ as defined in Eq. (\ref{anodime}), {\it vs.} $t$. Same notation as in Fig. \ref{figr3}. }
\label{figr4}\end{center}
\end{figure}

The anomalous dimension $\eta$, computed  from the field renormalization as
\begin{equation}
 \eta=\partial_t {\rm ln} (W_\pi) \, ,
\label{anodime}
\end{equation} 
is reported in Fig. \ref{figr4} for the same initial values $\underline \alf^2$ and with the same notation
used in Fig. \ref{figr3}.  In the presence of a fixed point, $\eta$ must approach asymptotically a constant value, as it is almost realized in the first group of curves.  
Even in this figure the dissimilar regimes of the  three groups are evident.

\begin{figure}[tb]
\begin{center}
\includegraphics[width=8cm]{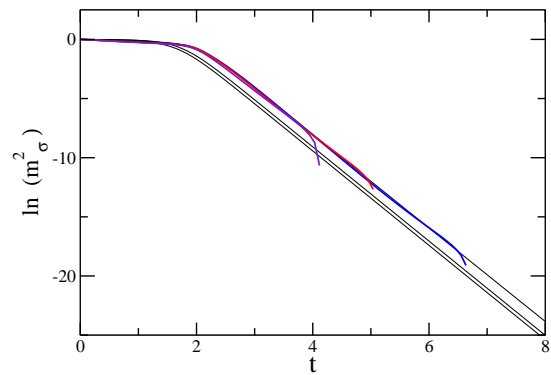}
\caption{ Flow of
${\rm ln}(m_\sg^2)$ for the following cases (from bottom to top) : $\underline \alf^2$ = 1.5,  0.8,  0.3 (solid, black online),
and 0.23,  0.22, 0.2 (red, blue and violet online).}
\label{figr5}\end{center}
\end{figure}
\begin{figure}[tb]
\begin{center}
\includegraphics[width=8cm]{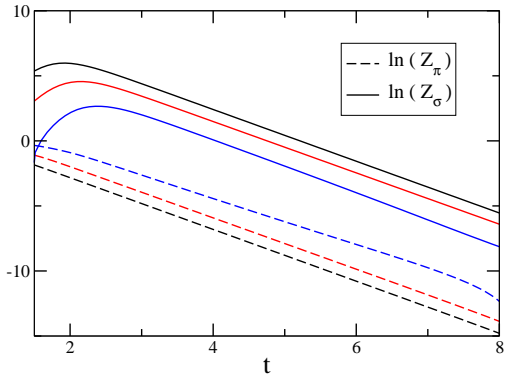}
\caption{Flow of  ${\rm ln}( Z_\pi)$ (dashed)  and  ${\rm ln}( Z_\sg)$ (solid) for 
$\underline \alf^2$ =1.1 (external, black online curves), 0.5 (intermediate, red online), 0.24 (inner, blue online).}
\label{figr6}\end{center}
\end{figure}

As already noticed, the parameters  $m_\sg^2$, $Z_\pi$  and $Z_\sg$ are expected to be relevant and therefore an accurate adjustment of their initial values is required
to avoid a rapid breakdown of the flow of the various parameters and so, for instance, in the case with $\underline\alf^2=0.2$ we found  $\underline m_\sg^2=0.0175$ and 
$\underline Z_\pi=\underline Z_\sg =-8.665\, 10^{-3}$, while, with $\underline\alf^2=0.8$, we got  $\underline m_\sg^2=0.498$ and essentially the same value of the previous case for the other two parameters.
The quantity ${\rm ln}(m_\sg^2)$ (with $m_\sg^2$ normalized to its initial value) is shown in  Fig. \ref{figr5} for the following cases: $\underline \alf^2= 1.5, \, 0.8, \, 0.3$ (the three curves that reach $t=8$)
and  $\underline\alf^2=  0.23, \, 0.22,\, 0.2$ (the remaining three curves that rapidly turn downwards at  $t < 8$). After the initial transient interval, the curves of the first group  become straight lines 
with common slope, which is in agreement with the power law dependence, $m_\sg^2\propto k^{4-\eta}$, characteristic of the scaling at the LP.

In the same spirit of Fig. \ref{figr5},  we show in Fig. \ref{figr6}  the logarithm of $Z_\pi$  (dashed) and $Z_\sg$  (solid) in the straight line regime, for $\underline\alf^2= 1.1$ (most external curves, black online),
$0.5$  (intermediate curves, red online), and  $0.24$ (inner curves, blue online).  Apart from the last case where some deformation shows up, in the other two cases the slope of  
$Z_\pi$  and $Z_\sg$ is essentially the same and corresponds to the expected exponent:  $Z_\pi\propto k^{2-\eta}$.

%%%%%%%%%%%%%%%%%%%%%%%%%%%%%%%%%%%%%%%%%%%%%%%%%%%%%%%%%%%%%%%%%%%%%%%%%%%%%%
%%%%%%%%%%%%%%%%%%%%%%%%%%%%%%%%%%%%%%%%%%%%%%%%%%%%%%%%%%%%%%%%%%%%%%%%%%%%%%
\vskip 8 pt
{\section{Conclusion}
\label{VI}}
Prompted by the similarities  observed between the characteristics of the Wilson-Fisher fixed point in $d/2$ dimension and those of the $d$-dimensional isotropic LP,    
in this paper we searched for indications  of a continuous line of LP for the  $U(1)$ symmetric scalar theory at  the isotropic LP lower critical dimension, $d=4$, 
analogous to the line of fixed points observed in $d=2$. The latter fixed point  line, which cannot be related to a standard order-disorder phase transition because of the Coleman-Mermin-Wagner 
theorem, is instead associated to the BKT transition, which is a transition of topological nature to a  quasi-ordered phase exhibiting  the vanishing of the order parameter
and an algebraic  decay of the correlator.  

In particular, we limited our analysis  to the determination of LP solutions and to the study of  their scaling properties, without investigating on the explicit, presumably topological, nature of the mechanism  that drives 
the transition. To this aim, we exploited the analysis presented in \cite{jame} where, with the help of a particular approximation of the FRG flow equations, many peculiar properties  of BKT phase were recovered;
therefore, we considered a suitable truncation in the expansion of the effective action in powers of the field and solved the corresponding flow equations in the form of a set of  ordinary differential equations.

Whereas for the BKT phase  the algebraic decay of the order parameter correlations is observed by using a modulus-phase representation of the complex field $\phi$ and within a minimal truncation 
of the derivative sector of the effective action, that includes only two parameters to $O(\partial^2)$ (in addition to the effective potential approximated to the fourth power of $\phi$),  in the case of the LP we found that
the same  result is not  recovered for the equivalent minimal truncation.  However,  this minimal truncation does not correspond to a systematic truncation in the derivative  expansion for the LP case. 
Then, by systematically including  all operators  up  to $O(\partial^4)$ and, at the same time, up to $\phi^4$  in the ansatz of the effective action, we observed that it is possible to find suitable combinations
of the coefficients of the various gradient terms, that reduces the effective action to a quadratic action in the angular field $\vartheta$, through suitable cancellations. This immediately leads to the searched algebraic 
decay. Moreover, the fixed point nature of the quadratic effective action should protect against possible changes of the coefficients due to infrared fluctuations, that could spoil the cancellation of the unwanted terms.

Then, turning to the usual cartesian representation of $\phi$ in terms of real longitudinal and transverse  components, we analyzed the flow of the parameters corresponding to the minimal truncation 
of the effective action studied in \cite{jame}. In this case, as observed for the BKT phase in \cite{jame}, we found that the suppression of the longitudinal fluctuations leads to a very clear picture of a line of  LPs,
parametrized by the stiffness. In addition, we studied the spectrum of eigenvalues for the perturbations around these LPs, pointing out the presence of a relevant operator associated to the 
$O(\partial^2)$ term of the transverse field. As expected, this relevant direction, not present in the  BKT phase, is now generated because the leading derivative operator  in the LP case 
corresponds to  the  $O(\partial^4)$ term, whose scaling is regulated by the anomalous dimension only. Consequently, due to dimensional arguments, the general  scaling at a LP requires 
the $O(\partial^2)$ term to be relevant.

Finally, we  included  the longitudinal fluctuations and considered the resulting FRG equations that, similarly to the BKT case,  no longer exhibit an exact line of fixed points, but rather produce a flow which, at least for very large 
initial values of the stiffness $J$, shows the typical  scaling of the parameters at a LP for a very long interval of the running parameter $t$. However, in this approximation  the exact fixed point solution is missing 
and the regime of the flow described above eventually breaks down at some extremely large value of $t$. At smaller $J$, the interval of $t$ where the quasi-fixed point regime is realized  gets smaller 
and the breakdown scale becomes  visible; at even smaller $J$, it disappears essentially in the same way as it is found in \cite{jame} for the BKT phase.

We conclude that the particular approximation scheme adopted here for the LP in $d=4$, substantially  reproduce the same properties that are observed, within the same approximation, in the BKT phase of the $U(1)$ model 
in $d=2$. In the latter case, the reliability of the approximation depends on the fact that the results obtained, although not accurate in the region of small $J$, for large $J$ reproduce specific properties of the
BKT phase, that are established otherwise. Turning to our four-dimensional problem, we cannot count on alternative evidences of a line of LPs that possibly ends at  a transition point to a new phase; therefore, we 
take the findings obtained in this approximation as a first indication of a non-trivial aspect of the  $U(1)$ theory, associated to the Lifshitz scaling and possibly related to a transition of topological nature.

%%%%%%%%%%%%%%%%%%%%%%%%%%%%%%%%%%%%%%%%%%%%%%%%%%%%%%%%%%%%%%%%%%%%%%%

\vskip 8 pt

\begin{acknowledgments}
The author is grateful to M. Shpot and P. Jakubczyk for discussions and valuable suggestions. This work has been carried out within the INFN project QFT- HEP.
\end{acknowledgments}

%%%%%%%%%%%%%%%%%%%%%%%%%%%%%%%%%%%%%%%%%%%%%%%%%%%%%%%%%%%%%%%%%%%%%%%


\begin{thebibliography}{99}

\bibitem{Horn} R.M. Hornreich, M. Luban, S. Shtrikman, Phys. Rev. Lett. {\bf 35}, 1678, (1975).

\bibitem{erzan} A. Erzan, G. Stell, Phys. Rev. {\bf B 16}, 4146, (1977).

\bibitem{sak} J. Sak,  G.S. Grest, Phys. Rev. {\bf B 17},  3602, (1978). 

\bibitem{grest} G.S. Grest, J. Sak, Phys. Rev. {\bf B 17}, 3607, (1978).

\bibitem{Diehl}  H.W. Diehl,  Acta Phys.Slov. {\bf 52}, 271, (2002).


\bibitem{selke1988}  W. Selke, Phys. Rep. {\bf 170}, 213, (1988).

\bibitem{horava}  P. Horava, Phys. Rev. {\bf D79}, 084008, (2009).
e-print arXiv: 0901.3775 [hep-th].

\bibitem{filippo}  D. Benedetti,  F. Guarnieri, Journal of High Energy Physics {\bf 1403}, 078, (2014).
e-print arXiv: 1311.6253 [hep-th].

\bibitem{zz}  G. D'Odorico, F. Saueressig, M. Schutten, Phys. Rev. Lett. {\bf 113}, 171101,  (2014). 
e-print arXiv: 1406.4366 [gr-qc].

\bibitem{cognola} G. Cognola, R. Myrzakulov, L. Sebastiani , S. Vagnozzi, S. Zerbini,  Class. Quant. Grav. {\bf 33},  225014,  (2016).
e-print arXiv: 1601.00102 [gr-qc].

\bibitem{bekaert1} X. Bekaert, M. Grigoriev, Nucl. Phys. {\bf B876},  667, (2013).
e-print arXiv: 1305.0162 [hep-th].

\bibitem{bekaert2}
X. Bekaert, M. Grigoriev,  Bulg. J. Phys. {\bf 41}, 172, (2014).

\bibitem{Alexandre} J. Alexandre, Int. J. Mod. Phys. {\bf A26}, 4523, (2011).
e-print arXiv: 1109.5629 [hep-ph].

\bibitem{kikuchi} K. Kikuchi, Prog.Theor.Phys. {\bf 127},  409, (2012).
e-print arXiv: 1111.6075 [hep-th].

\bibitem{casal2} R. Anglani, R. Casalbuoni, M. Ciminale, N. Ippolito, R. Gatto, M. Mannarelli, M. Ruggieri,  Rev. Mod. Phys.  {\bf 86}, 509, (2014). 
e-print arXiv: 1302.4264 [hep-ph].

\bibitem{buballa}
M.  Buballa, S. Carignano,  Prog. Part. Nucl. Phys. {\bf 81}, 39, (2015).
e-print arXiv: 1406.1367 [hep-ph].

\bibitem{pisarski} R.D. Pisarski, V. V. Skokov, A.M. Tsvelik, {\it Anisotropic fluctuations in cool quark matter and the phase diagram of Quantum Chromodynamics }.
e-Print  arXiv: 1801.08156 [hep-ph].


\bibitem{parisshpot} R.B. Paris, M.A. Shpot, Math. Methods  Appl. Sci. {\bf 41}, 2220, (2018).
e-Print arXiv: 1707.03018 [hep-th].

\bibitem{carneiro} C. Mergulh\~ao Jr.,  C.E.I. Carneiro,  Phys. Rev. {B 59}, 13954, (1999).

\bibitem{diehl1} H. Diehl, M. Shpot,  Phys. Rev. {\bf B 62}, 12338, (2000).
e-print arXiv: cond-mat/0006007.

\bibitem{diehl2} M. Shpot,  H. Diehl,  Nucl. Phys. {\bf B 612}, 340, (2001).

\bibitem{diehl3} M.A. Shpot, H.W. Diehl, Y.M. PisÕmak,   J. Phys.  {\bf A  41}, 135003, (2008).
e-print  arXiv: 0802.2434  [cond-mat.stat-mech].

\bibitem{Wetterich:1992yh} C. Wetterich,  Phys. Lett.  {\bf B 301}, 90, (1993).

\bibitem{Morris:1994ie} T.R. Morris, Int. J.  Mod. Phys. {\bf A9}, 2411, (1994).
e-print arXiv: hep-ph/9308265.
  
\bibitem{Berges:2000ew} J. Berges, N. Tetradis, C. Wetterich, Phys. Rept.  {\bf 363}, 223-386, (2002).
e-print arXiv: hep-ph/0005122.

\bibitem{bervillier} C. Bervillier, Phys. Lett. {\bf A331},110, (2004).
e-print arXiv: hep-th/0405027.

\bibitem{Essafi} K. Essafi, J.P. Kownacki, D. Mouhanna, Europhys.Lett. {\bf 98},  51002, (2012).
e-print  arXiv: 1202.5946 [cond-mat.stat-mech].

\bibitem{diehl4} H.W. Diehl, M.A. Shpot, Journal of Physics. {\bf A 35}, 6249, (2002).
e-print arXiv: cond-mat/0204267.

\bibitem{schmid}
M. Mueller, F. Schmid,
ADVANCED COMPUTER SIMULATION APPROACHES FOR SOFT MATTER SCIENCES II - Book Series: Advances in Polymer Science,  {\bf 185}, 1,  (2005). 
e-print arXiv: cond-mat/0501076.


\bibitem{boza} A. Bonanno,  D. Zappal\`a,  Nucl. Phys. {\bf B893}, 50, (2015).
e-print  arXiv: 1412.7046 [hep-th].

\bibitem{zappa} D. Zappal\`a,  Phys. Lett. {\bf B773}, 213, (2017).
e-print arXiv: 1703.00791 [hep-th] .

\bibitem{liao} S.-B. Liao, Phys. Rev. {\bf D53}, 2020, (1996).
e-print arXiv: hep-th/9501124.

\bibitem{bohr} O. Bohr, B.-J. Schaefer, J. Wambach, Int. J. Mod. Phys. {\bf A16}, 3823, (2001).
e-print arXiv: hep-ph/0007098.
      
\bibitem{boza2001} A. Bonanno,  D. Zappal\`a,  Physics Letters {\bf B504}, 181, (2001).
e-print arXiv: hep-th/0010095.

\bibitem{Litpaw3} D.F. Litim, J.M. Pawlowski, Phys. Rev. {\bf D66}, 025030, (2002).
arXiv: hep-th/0202188.

\bibitem{Litpaw4}
D.F. Litim, J.M. Pawlowski, Phys. Lett. {\bf B546}, (2002), 279-286.
e-print arXiv:  hep-th/0208216.

\bibitem{zap2001}  D. Zappal\`a, Physics Letters {\bf A290}, 35, (2001).
e-print arXiv: quant-ph/0108019.

\bibitem{Maza} M. Mazza, D. Zappal{\`a}, Phys. Rev. {\bf D64}, 105013,  (2001).
e-print arXiv: hep-th/0106230.
      
\bibitem{litimzappala} D.F. Litim, D. Zappal\`a,  Phys. Rev. {\bf D83}, 085009, (2011).
e-print arXiv: 1009.1948 [hep-th].

\bibitem{safari1} M. Safari, G.P. Vacca,  Phys. Rev. {\bf D 97}, 041701, (2018).
e-Print arXiv:1708.09795 [hep-th].

\bibitem{safari2} M. Safari, G.P. Vacca, Eur. Phys. J. {\bf C78}, 251, (2018).
e-Print arXiv:1711.08685 [hep-th].

\bibitem{gubser} S.S. Gubser, C. Jepsen, S. Parikh, B. Trundy, Journal of High Energy Physics  {\bf 1711}, 107, (2017).
e-Print arXiv:1703.04202 [hep-th].

\bibitem{mermin66} N.D. Mermin, H. Wagner, 
Phys. Rev. Lett. {\bf 17}, 1133, (1966).

\bibitem{coleman} S. Coleman,  Commun. Math. Phys. {\bf 31}, 259, (1973).

\bibitem{berezinskii71} V.L. Berezinskii, 
Sov. Phys. JETP {\bf 32}, 493, (1971).

\bibitem{kosterlitz73} J.M. Kosterlitz,  D.J. Thouless, 
J. Phys. {\bf C 6}, 1181, (1973).

\bibitem{grater} M. Gr\"ater, C. Wetterich, 
Phys. Rev. Lett. {\bf 75}, 378, (1995).
e-print arXiv: hep-ph/9409459.


\bibitem{wetterich} G.v. Gersdorff,  C. Wetterich,  
Phys. Rev.  {\bf B 64}, 054513, (2001).
e-print arXiv: hep-th/0008114.

\bibitem{dupuis} P. Jakubczyk, N. Dupuis, B. Delamotte, 
Phys. Rev.  {\bf E 90}, 062105, (2014).
e-print arXiv: 1409.1374 [cond-mat.stat-mech].


\bibitem{jame} P. Jakubczyk, W. Metzner,  
Phys. Rev.  {\bf B 95}, 085113, (2017).
e-Print arXiv:1606.04547[cond-mat.stat-mech].

\bibitem{popov87} V. N. Popov,  {\it Functional integrals and collective excitations},  (Cambridge University Press, Cambridge, 1987).

\bibitem{metz} B. Obert, C. Husemann, W. Metzner, 
Phys. Rev.  {\bf B 88}, 144508, (2013).

\end{thebibliography}
\end{document}